# Population spatialization and synthesis with open data

Ying Long, Beijing Institute of City Planning, China

Zhenjiang Shen, Kanazawa University, Japan

Abstract: Individuals together with their locations & attributes are essential to feed micro-level applied urban models (for example, spatial micro-simulation and agent-based modeling) for policy evaluation. Existed studies on population spatialization and population synthesis are generally separated. In developing countries like China, population distribution in a fine scale, as the input for population synthesis, is not universally available. With the open-government initiatives in China and the emerging Web 2.0 techniques, more and more open data are becoming achievable. In this paper, we propose an automatic process using open data for population spatialization and synthesis. Specifically, the road network in OpenStreetMap is used to identify and delineate parcel geometries, while crowd-sourced POIs are gathered to infer urban parcels with a vector cellular automata model. Housing-related online Check-in records are then applied to distinguish residential parcels from all of the identified urban parcels. Finally the published census data, in which the sub-district level of attributes distribution and relationships are available, is used for synthesizing population attributes with a previously developed tool Agenter (Long and Shen, 2013). The results are validated with ground truth manually-prepared dataset by planners from Beijing Institute of City Planning.

Acknowledgments: We acknowledge the financial support of the National Natural Science Foundation of China (No. 51408039). We thank Ms Liqun Chen for her proofreading.

## 1 Introduction

Spatial distribution of population individuals and their socioeconomic attributes are essential to feed micro-level applied urban models (such as spatial micro-simulation and agent-based modeling) for policy evaluation. A number of studies have been conducted to generate synthetic individual data by reweighting large-scale surveys (Wu et al, 2008). In developing countries like China, population distribution in a fine scale, as the input for population synthesis, is not universally available. Neither are large-scale surveys for population synthesis. China is also facing an institution-induced digital divide: a gap between data availability and official open datasets, as the government exercises tight control of official data. This situation is common in developing countries like China, Southeast Asian countries, South American countries, and African countries (Tatem and Linard, 2011). We aim to mitigate this gap by illustrating how the collection, analysis, and visualization of big (open) data can open up new avenues for synthesizing micro data in developing countries. In this paper, we prepare an alternative solution for population spatialization and synthesis simultaneously using open data, in the condition of without high-resolution population distribution and large-scale surveys.

The existed studies on population spatialization and synthesis are generally separated. There has been extensive research on mapping population distribution (Langford and Unwin, 1994; Mennis, 2003; Liao et al., 2010). The most common method used is interpolating population density with spatial factors and population census data. Most of the spatialized products, from statistic aggregated in administrative units (Linard et al, 2012) or nighttime satellite imagery (Sutton, 1997; Lo, 2001), are associated with a coarse resolution scale, which is not fine enough for micro-level urban models (for instance, 1-km population grids for China by CAS, 5km-grid UNEP/GRID for several continents, 2.5' grids GPW/GRUMP for the world, 1km-grid LandScan for the world. With the emerging techniques of high-resolution remote sensing images and datasets for individual buildings, several studies generate population distribution in a fine scale, e.g. 100 m or building level (Silvan et al, 2010; Azar et al, 2013; Silva et al, 2013). However, these methods could be time-consuming, expensive, and labor-intensive (Erickson et al 2013), which makes it difficult for many developing countries to conduct such studies. Most of the studies are for population at nighttime, with an exception of Bhaduri et al (2007) for both daytime and nighttime population distribution. In addition, these studies did not consider the disaggregation of population attributes.

Population synthesizing techniques have been developed for disaggregating population attributes, based on known population distribution. Reweighting and synthetic construction are the two dominant approaches for population synthesis (Hermes and Poulsen, 2012). Müller and Axhausen reviewed a list of population synthesizers, including PopSynWin, ILUTE, FSUMTS, CEMDAP, ALBATROSS, and PopGen (2010). Iterative proportional fitting (IPF) adopted by PopGen, a typical reweighting method, can adjust tables of data cells so they add up to selected totals for both the columns and rows (in two-dimensional cases). The unadjusted data cells are referred as seed cells, and the selected totals are referred as marginal totals. Synthetic construction can generate micro data with only aggregated information. This approach does not require individual samples. For instance, Barthelemy and Toint (2012) produced a synthetic population for Belgium at the municipality level without a sample. Long and Shen (2013) synthesized individuals with aggregated data, empirical studies and common sense for Beijing. Therefore, the approach of synthetic construction is more appropriate for applications in data-sparse developing countries.

Now, the emerging trend of open and big data provides opportunities for population spatialization and synthesis in developing countries. As one of the most successful volunteered GIS projects, OpenStreetMap (OSM) in developing countries has been encouraging, as the volume of OSM data in China has experienced a ninefold increase during 2007-2013 (Long and Liu, 2013). OSM has been proposed as a promising candidate for a quick and robust delineation of parcels (Haklay and Weber 2008), thus providing basic spatial units for allocating population at a fine scale. Points of interest (POIs) available in most of online mapping services are also promising for identifying residence related places (Long and Liu, 2013). The coupling of OSM and POIs would be an alternative for mapping population with a high-resolution. In addition, as we have reviewed, the sub-district census data can be used to feed synthetic construction method for population synthesis.

In this paper, we propose an automatic process using open data for population spatialization and synthesis. Specifically, the road network in OpenStreetMap is used to identify and delineate parcel geometries, while crowd-sourced POIs are gathered to infer urban parcels with a vector cellular automata model referring to our previous study (Long and Shen, 2014). Housing-related online Check-in records or POIs are then applied for selecting residential parcels from all identified urban parcels. Finally the published sub-district level population census, in which distribution of attributes and relationships between attributes are available, is used for synthesizing population attributes supported by a previously developed tool Agenter (Long and Shen, 2013). We focus on urban residents (spatial distribution at night time) in this study and rural ones would be reserved in our future research, since urban residents are the majority of the population in Beijing. We expect to feed applied urban models with the results in Beijing. In this paper, the study area and data are described in Section 2. The applied methodologies are elaborated in Section 3. We discuss the results and validations in Section 4 and 5, respectively. The concluding remarks are in Sections 6.

## 2 Study area and data

*2.1 Study area*

As the capital of China, the Beijing Metropolitan Area (BMA) with a coverage of 16,410 km$^2$ has over 20 million residents in 2010 and is becoming one of the most populous cities in the world. The BMA lies in northern China, to the east of the Shanxi altiplano and south of the Inner Mongolian altiplano. The southeastern part of the BMA is a plain, extending eastward for 150 km to the Bohai Sea. Mountains cover an area of 10,072 km2, 61% of the whole study area (Figure 1). See Yang et al. (2011) for more background information on Beijing.

According to Beijing Municipal Bureau of Statistics and NBS Survey Office in Beijing (2013), the total urban residents of BMA in 2012 were 17.837 million. According to the land use dataset of Beijing Institute of City Planning, the total urban area of BMA in 2012 was 1674.9 km$^2$, with 419.2 residential areas.

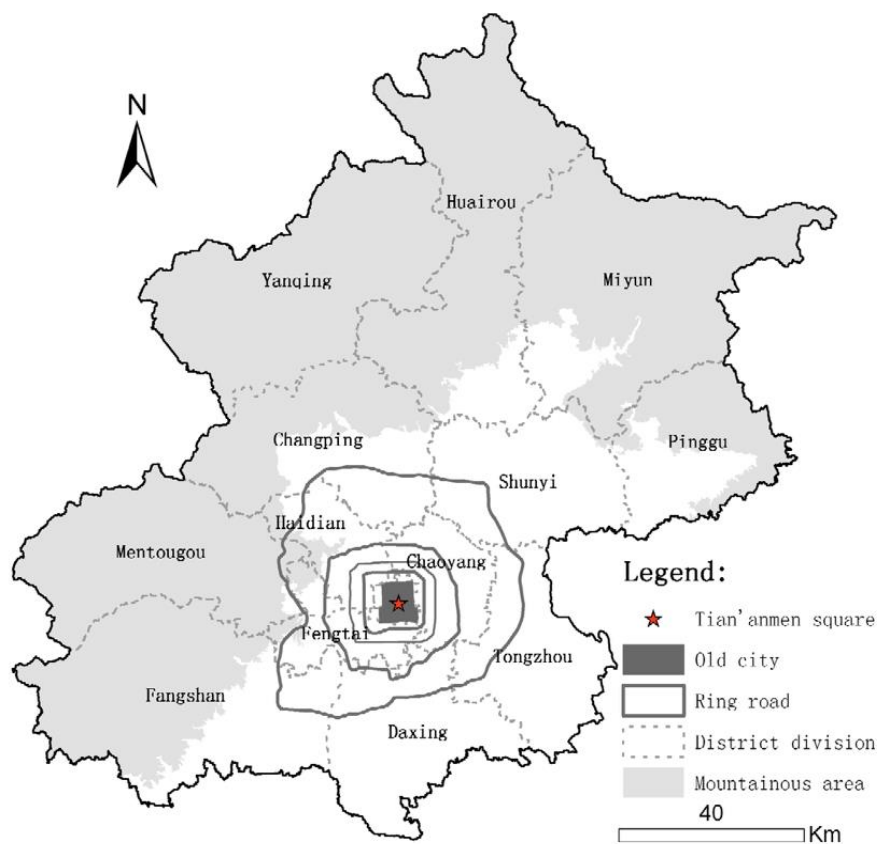

**Fig. 1.** The Beijing Metropolitan Area

*2.2 The OSM road networks of Beijing*

We downloaded OSM road networks for Beijing on October 5, 2013. The OSM dataset contains 43,006 road segments, in a total of 20,904 kilometers (Figure 2). We also collected the 2012 ordnance survey map of Beijing with detailed road networks to verify the results produced by OSM data. OSM data quality is promising, especially in large cities like Beijing as we found in 2013 (Long and Liu,2013).

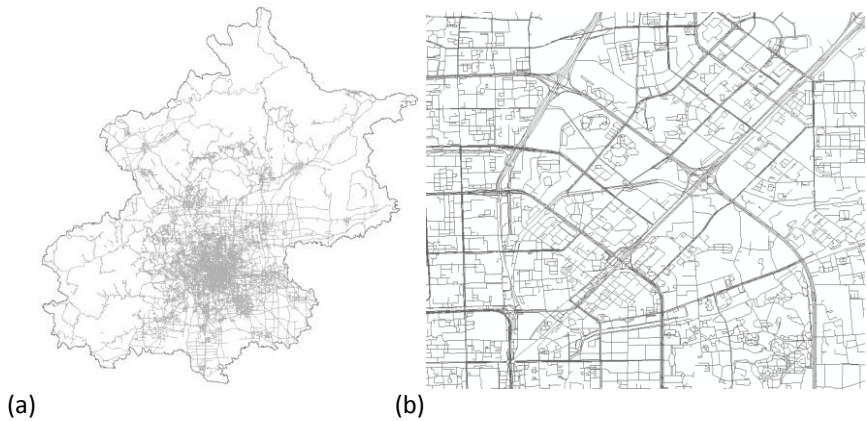

(a)         (b)

**Fig. 2.** The OSM road network of Beijing in 2013. (a) the whole BMA; (b) a part of the central city of Beijing

The city center and main roads are also extracted from the OSM data.We will use them as the spatial factors to identify urban parcels for mapping population.

## *2.3 POIs*

A total of 186,437 POIs are gathered and geo-coded by business cataloguing websites, among which 31,189 are residential POIs.The initial twenty POI types are aggregated into eight more general assemblies, including residential communities (RES), commercial sites (COM), business establishments (FIR), transportation facilities (TRA), government buildings (GOV), educational sites (EDU), green space (GRE) and others (OTH). The quality of data is guaranteed through manually checking on the randomly selected POIs. The total amount of POIs associated with each parcel is used for inferring the parcel's density and the residential POIs amount for mapping population.

It is worth noting that the POI counts can be replaced by other measures human activity, ranging from the conventional remoted sensing based land use cover to ubiquitous online check-in data (e.g., Foursquare) in the background of web 2.0.

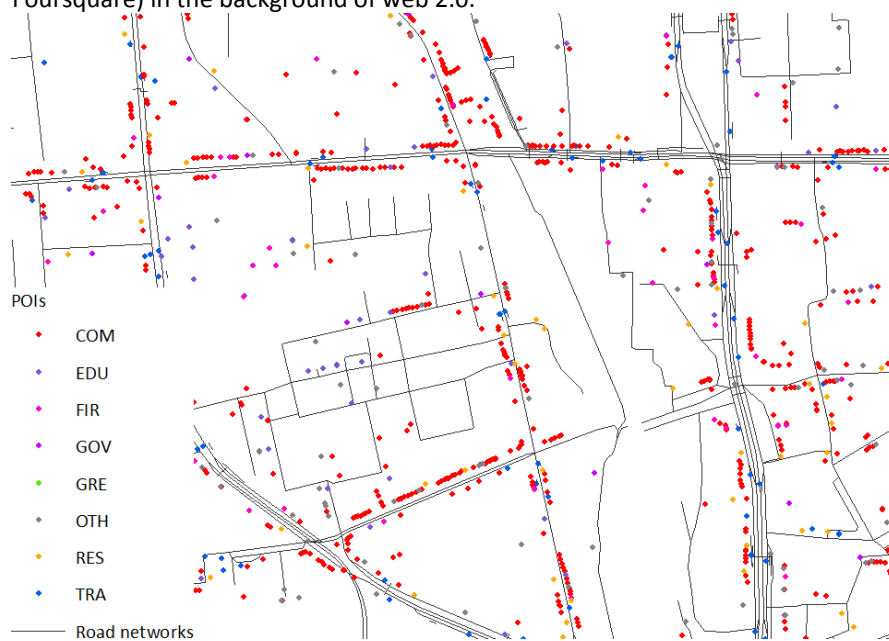

**Fig. 3.** Spatial distribution of POIs in Beijing by type (in a local area of Beijing)

## *2.4 The 2010 population census of Beijing*

The main data source for population synthesis in Beijing is the Sixth Population Census Report of the BMA conducted in 2010 (the census, detailed at http://www.bjstats.gov.cn/rkpc_6/pcsj/201105/t20110506_201581.htm). The census was conducted at the census tract level[1], which size is between a city block and sub-district. These data were aggregated from the original census tract level to the sub-district level (BMA has 307 sub-districts). The census data provide both the statistical distributions of attributes and relationships among attributes of residents. Many cross-tabulations for various combinations of the attributes are listed in the census report, and were used in this (Table 1). These are the attributes and locations of residential agents to be disaggregated within the BMA. The dependent relationships among attributes are illustrated in Figure 4. Note that we are about to spatialize and synthesize urban residents of Beijing in 2012, and there is a mismatch between the target year and the census year.

The resident count is from the statistical yearbook, and data only used for synthesizing population attributes.

**Table 1.** Descriptions and known information for each attribute of residential agents in the BMA

| Name | Description | Type | Known information | Data source | Data type | Order |
|---|---|---|---|---|---|---|
| AGE | Age in years | Non-spatial attribute | Frequencies | The census | Ratio | 1 |
| SEX | Gender | Non-spatial attribute | Frequencies | The census | Nominal (male, female) | 2 |
| MARRIAGE | Marital status | Non-spatial attribute | Frequencies, RB (with AGE) | The census | Nominal (married, unmarried, divorced, remarried, widowed) | 3 |
| EDUCATION | Level of education | Non-spatial attribute | Frequencies, RB (with AGE) | The census | Ordinal (junior middle school, undergraduate, etc.) | 4 |
| JOB | Occupation | Non-spatial attribute | Frequencies, RB (with EDUCATION) | The census | Nominal | 5 |
| INCOME | Monthly income | Non-spatial attribute | Frequencies | The survey | Ratio | 6 |
| FAMILYN | Number of family members | Non-spatial attribute | Frequencies | The census | Ordinal (one person, two person, etc.) | 7 |
| PARCEL | ID of parcel at which the agent resides | Location | Frequencies | An empirical study | Nominal | 8 |
| TAM | Distance to the city center | Spatial attribute | Location of Tiananmen Square | Urban GIS | Ratio | 9 |

Note: The attribute PARCEL is the ID of the parcel used to map the disaggregated residents.

---

[1] The spatial distribution of census tracts has never been released by the Beijing Municipal Statistical Bureau. Therefore, it is not possible to determine whether census tracts are compatible with TAZs.

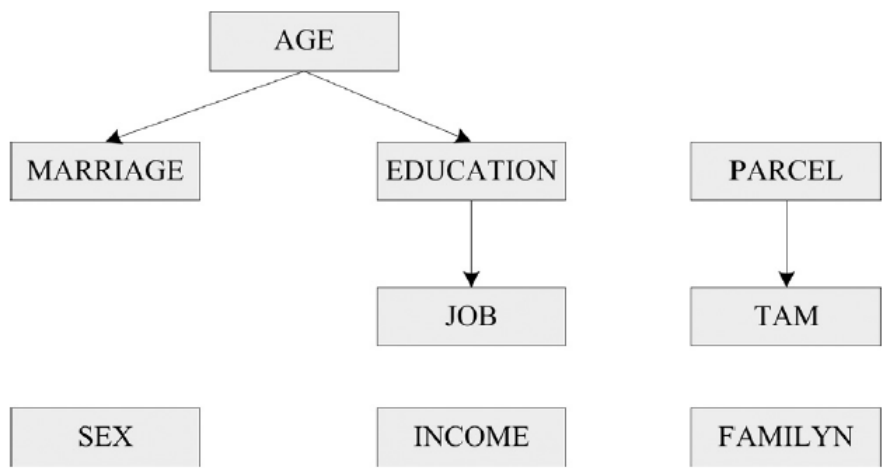

**Fig. 4.** Dependent relationships among attributes of residential agents.

## 3 Approach

### 3.1 The proposed framework

The framework for population spatialization and synthesis is as follow: (1) Generating parcels using road networks (Section 3.2); (2) Selecting urban parcels using POIs and constraints with vector cellular automata (Section 3.3); (3) Identifying residential parcels from all urban parcels selected using residential POIs (Section 3.4); (4) Allocating urban population into identified residential parcels using totals in the sub-district level and residential POI density (Section 3.5); (5) Synthesizing urban population attributes using the census and empirical research with Agenter (Section 3.6).

### 3.2 Generating parcels

The working definition of a parcel is a continuously built-up area bounded by roads. Identifying land parcels and delineating road space are therefore *dual* problems. In other words, our approach begins with the delineation of road space, and individual parcels are formed as polygons bounded by roads.

The delineation of road space and parcels is performed as follows: (1) All OSM road data are merged as line features in a single data layer; (2) individual road segments are trimmed with a threshold of 200m to remove hanging segments; (3) individual road segments are then extended on both ends for 20m to connect adjacent but non-connected lines; (4) road space is generated as buffer zones around road networks. A varying threshold ranging between 2-30m is adopted for different road types (e.g. surface condition, as well as different levels of roads); (5) parcels are delineated as the space left when road space is removed; and (6) a final step involving overlaying parcel polygons with administrative boundaries to determine whether individual parcels belong to a certain administrative unit. Parameters used in these steps are determined pragmatically with topological errors of OSM data in mind.

Four parameters are further calculated for each parcel. The first two, size and compactness, are determined by the geometric characteristics of each parcel. The third, the accessibility, is taken into consideration as a location variable for describing a parcel. The last one is the functional attribute of a parcel for reflecting its actual use. POIs within or close to a parcel are measured as its urban density. Due to the natural unevenness of urban density between central cities and other ones, POIs density is further normalized and placed between 0 and 1 to release the heterogeneity among cities. Because of lacking further attributes of POIs, the popularity of POIs is assumed as the same in this study. When any substitutions are available, they can be expected to approximate the intensity of urban activities explicitly. Hence, in the way of using road network and POIs to identify initial parcels, the spatial and functional features are incorporated together for further urban parcel selection.

### 3.3 Selecting urban parcels

Vector-based constrained cellular automata models are used for picking up urban parcels from the initial ones generated by road network in diverse cities. We suppose this process is similar to that for modelling urban expansion, which sees extensively CA applications. Apart from the conventional raster CA model (Batty, et al, 1999), vector-based CA model here depends on irregular polygons rather than regular cells. In this research, each parcel is regarded as a cell with a status that is 0 (urban) or 1 (non-urban). This can be illustrated as a formula as the following.

$$S_{ij}^{t+1} = f(S_{ij}^{t}, \Omega_{ij}^{t}, Con, N)$$

Here, a parcel's status at t +1 is considered as a function *f* of parcel's statues and other factors at *t*. In this function, $S_{ij}^{t}$ and $S_{ij}^{t+1}$ denote to the statues of parcels at time points of *t* and *t+1* respectively; $\Omega_{ij}^{t}$ is the neighboring situation; *Con* refers to the constraints and *N* is the amount of all parcels. This function can be further transferred to a detailed probability formula:

$$P_{ij}^{t} = (P_{l})_{ij} \times (P_{\Omega})_{ij} \times con(\cdot) \times P_{r}$$

In this function, the possibility of transformation of parcel's state at *t* is illustrated as multiplied product of probabilities of factors. Specifically, $(P_{l})_{ij}$ stands for the local potential that a parcel would convert its status from the non-urban to the urban while $(P_{\Omega})_{ij}$ denotes the conversion possibility in terms of the neighboring situations; $con(\cdot)$ stands for constraints and $P_{r}$ is the stochastic term.

The proposed spatial and functional characteristics are reflected in measuring the local potential. This could be explained in the formula below using a logistic regression model (Wu, 2002):

$$(P_{l})_{ij} = \frac{1}{1+\exp[-(a_0 + \sum_{k=1}^{m} a_k c_k)]}$$

where $a_0$ is a constant, $a_k$ is an estimated coefficient responding to the spatial variable $c_k$ and *m* is the total amount of spatial variables. As a result, spatial and functional factors are bonded to reflect the parcel's status in this study. Parcel size is measured in the natural logarithm of area. Compactness is calculated as perimeter square subdivided by area. Accessibility is abstracted as the minimum Euclidian distance to the city center. On the other hand, the functional factor is presented by applying the standardized POIs density, which is calculated as the rate of raw density in the max density among the samples.

The neighboring potential for a parcel is measured by the amount of peripheral urban parcels around it. This can be defined as:

$$(P_W)_{ij} = \frac{\sum con(S_{ij}^t = urban)}{n}$$

For parcel $ij$, $con(S_{ij}^t = urban)$ stands for the urban parcels within fixed areas while $n$ is the sum of all accessible parcels. The adjacent relation is defined as 500 m around the parcel $ij$.

Two layers - the steep area (a slope over 25 degrees) and various water bodies, are included as the constraints. Urban expansion is forbidden in these areas. The constraints are expressed as $con(cell_{ij}^t = suitable)$ with a value of 0 or 1, where 1 indicates that there is no restriction on the parcel's development as urban while 0 indicates that the parcel is forbidden to be urban.

The stochastic disturbance $P_r$ in the model stands for any possible change of local policies and accidental errors. It is calculated using

$$P_r = 1 + (-\ln\gamma)^\beta$$

where $\gamma$ is a random number ranging from 0 to 1, and $\beta$, ranging from 0 to 10, controls the effect of the stochastic factor.

Furthermore, by comparing the measured probability $(P_l)_{ij}$ with a fixed threshold value $P_{thd}$, the parcel's status at $t+1$ could be detected. If the measured value is greater than the threshold, the parcel is considered to be urban, if not, the parcel will stay as non-urban. This progress can also be presented as a binary expression:

$$S_{ij}^{t+1} = \begin{cases} Urban \text{ for } P_{ij}^t \succ P_{thd} \\ NonUrban \text{ for } P_{ij}^t \leq P_{thd} \end{cases}$$

Finally, for controlling the total area of all urban parcels, the urban area in 2012 is applied as the upper limits for the total area of selected urban parcels. is the area of parcel ij, and denotes to the reported total urban area for the BMA.

### 3.4 Identifying residential parcels

We define residential density as the ratio between the counts of residential POIs in/close to a parcel to the parcel area[2]. We further standardized the residential density to range from 0 to 1 using the following equation: standardized residential density = log(raw)/log(max), where raw and max correspond to density of individual parcels and the city-wide maximum residential density value[3]. We rank all selected urban parcels in terms of the calculated residential density. Residential parcels are identified via benchmarking the density with the residential area as the total control.

---

[2] Residential POIs within the buffered road space were accounted by their closest parcels in our experiment.

### *3.5 Allocating urban population*

For each identified residential parcel, we assume its population is proportional to its inferred residential POI density. In each sub-district of Beijing, we then allocate the total population into residential parcels.

### *3.6 Synthesizing population attributes using Agenter*

Population attributes are divided into two types, non-spatial attributes (such as age, income, and education for a residential agent) and spatial attributes (such as access to subways and amenities, land use, and height of the building that the residential agent occupies). The approach to disaggregating spatial attributes also differs from that used for non-spatial attributes.

The probability distribution of an attribute (hereafter referred to as the distribution) and the dependent relationship among attributes (hereafter referred to as the relationship) can be inferred from existing data sources, including aggregate data, small-scale surveys and empirical studies. Aggregate data include the total number, distribution and relationship (such as the cross-tabulation of marriage-age standing for the dependent relationship of marriage and age, and the cross-tabulation of income-education standing for the dependent relationship of income and education) of agents. Small-scale surveys that store samples can also be used to deduce the distribution of an attribute and the relationships among attributes. The probability distribution of an attribute and its relationship with other attributes can also be deduced using empirical studies. To convert aggregate data to individual samples, the probability distribution of the attributes and the relationship between them must be estimated. For more details on population synthesis, please refer to Long and Shen (2013).

### *3.7 Model validation*

According to our methods and research questions, validating our proposed framework includes two steps. The first is to validate population spatialization results and the second is to validate population synthesis results.

First, the selected residential parcels are compared with manually prepared data by planners in Beijing Institute of City Planning. The inferred population of each residential parcel is then correlated with total floor space of buildings within the parcel using building footprints and the floor number.

Second, the synthesized population of Beijing is validated by calculating the similarity between disaggregated and observed agents (using the 2010 Beijing household travel survey). We used the similarity indicator (SI) proposed by Long and Shen (2013) for comparison. The similarity index SI is 100% if two sets have the same agent attribute values. To calculate SI, both sets must be sorted by the same rule. The location attribute should be sorted first, followed by the other attributes in increasing order. It is also necessary to disaggregate the same number of agents as observed agents.

---

[3] The unit is the POI count per $km^2$. For parcels with no residential POIs, we assume a minimum density of 1 POI per $km^2$.

## 4 Results

### *4.1 Population spatialization*

We generated 52,359 parcels delineated by road networks in the BMA. To select urban parcels from all parcels, logistic regression is conducted for calibrating the weights for constraints in the proposed vector CA model. The 2010 parcel dataset in Beijing City, which is manually prepared by urban planners in BICP, is applied to mine rules for inferring residential parcels. It covers an area of 12,183 km$^2$ at a detailed urban parcel scale (Yanqing and Miyun counties in the Beijing Metropolitan Area are excluded from Beijing City). There are in total 52,330 parcels reported, among which 36,914 are identified as urban parcels.

According to the result of binary logistic regression (Table 2), 78.9% of all parcels can be explained by the generated function. And all factors except compactness have passed *p* test, revealing that they are significantly related to the differences between non-urban and urban ones. By using the logistic regression results, we selected 43,180 urban parcels. In order to test the accuracy of this model, the results of Beijing City with the CA model was compared with the BICP dataset again, and then an overall accuracy of 81.5% indicates the applicability of our model in delineating urban areas in terms of urban parcels.

**Table 2.** Binary logistic regression results for BICP parcels

| Name | Coefficient | S.E, | Sig. |
|---|---|---|---|
| Constant | 5.359 | .058 | 0.000 |
| Natural logarithm of parcel size | -0.306 | .006 | 0.000 |
| Distance to the city center | -0.099 | .001 | 0.000 |
| POIs density | 3.431 | .085 | 0.000 |

We then selected 7,161 residential parcels from all selected urban parcels and allocated urban residents on them (see Figure 5).

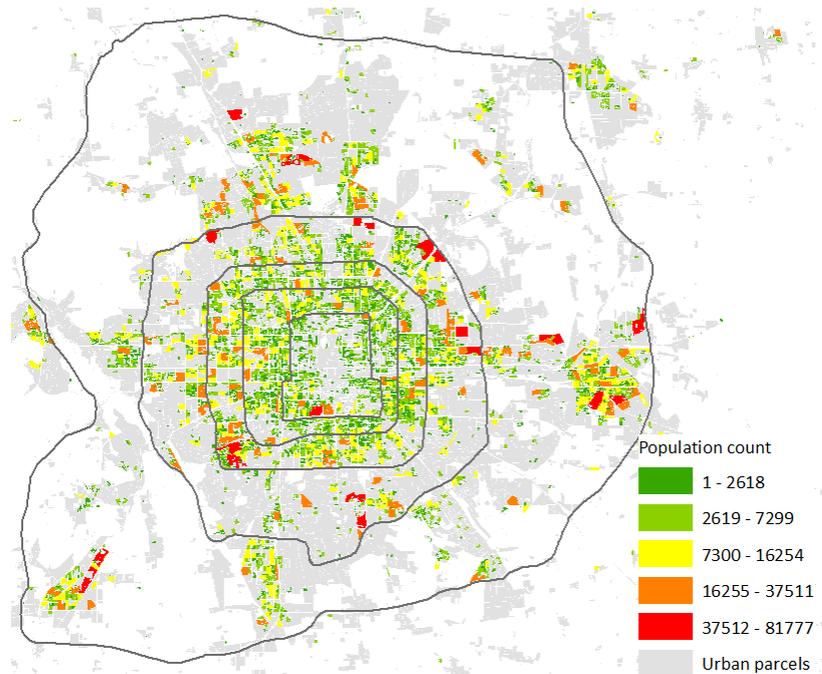

**Fig. 5.** Selected residential parcels and their population count

## *4.2 Population synthesis*

The synthesized urban residents are stored as the point Feature Class in the ESRI Personal Geodatabase (Table 3). This dataset embeds both the attributes and location information of residential agents, which can be regarded as a primary dataset for urban studies and initializing agent-based models.

**Table 3.** Disaggregated residential agents (partial) in the BMA.

| AID | AGE | SEX | MARRIAGE | EDUCATION | JOB | INCOME | FAMILYN | PARCEL | TAM |
|---|---|---|---|---|---|---|---|---|---|
| 193392 | 36 | Male | Married | Junior High/Middle School | Production, Transport Equipment Operator, and Related | 2385 | Three persons | 888 | 2140 |
| 198316 | 41 | Female | Married | High School | Production, Transport Equipment Operator, and Related | 5966 | Three persons | 966 | 7747 |
| 37094 | 61 | Male | Married | High School | Professional Technology Employee | 4744 | Three persons | 523 | 5721 |
| 165014 | 27 | Male | Unmarried | High School | Business and Service Employees | 5559 | Five persons | 768 | 4957 |
|  | 41 | Female | Married | Elementary School | Production, Transport Equipment Operator and Related | 5351 | Three persons | 18 | 36739 |
| 49808 | 21 | Male | Unmarried | Junior High/Middle School | Business and Service Employees | 2684 | Five persons | 274 | 2905 |
| 189128 | 21 | Male | Married | Junior High/Middle School | Production, Transport Equipment Operator and Related | 2578 | One person | 878 | 4092 |
| 118806 | 8 | Male | Unmarried | Elementary School | Production, Transport Equipment Operator and Related | 0 | Three persons | 478 | 6949 |
| 33570 | 53 | Female | Married | Elementary School | Production, Transport Equipment Operator and Related | 1304 | Five persons | 929 | 23760 |
| 179469 | 50 | Male | Married | Elementary School | Farming, Forestry, Animal Husbandry and Fishery | 4978 | Two persons | 804 | 2286 |

**5 Validation**

*5.1 Validating residential parcels with ground truth from BICP*

Table 4 shows a comparison between urban parcels generated by our approach and those contained in the BICP Beijing parcel data. It suggests that OSM-based approach generally produce larger parcels, due to the lack of information about tertiary and detailed roads in the OSM dataset[4]. Nevertheless, the two results are 71.2% matched with each other (the total area of intersected urban parcels in both data divided by the total area of OSM-based urban parcels), suggesting that they capture similar information on the geographic distribution of urban parcels and land use activities. In addition, we decompose the city of Beijing into sub-regions bounded by major ring roads, and calculate the proportion of parcels falling into individual sub-regions. The proportion of parcels falling into sub-regions between ring roads is consistent across both datasets.

**Table 4.** Comparison of selected urban parcels in BICP and OSM in Beijing (R=ring road)

| Parcels | Parcel count | Average size (ha) | Overlapped with BICP | Spatial distribution (in terms of area, km2) | | | | | |
|---|---|---|---|---|---|---|---|---|---|
| | | | | Within R2 | R2-R3 | R3-R4 | R4-R5 | R5-R6 | Beyond R6 |
| OSM | 7,130 | 17.2 | 1194.2 km2 (71.2%) | 42.5 | 74.0 | 113.4 | 263.5 | 666.5 | 519.9 |
| BICP | 57,818 | 2.9 | - | 48.6 | 69.7 | 99.8 | 229.5 | 687.9 | 544.4 |
| OSM/BICP | 0.12 | 5.93 | - | 0.87 | 1.06 | 1.14 | 1.15 | 0.97 | 0.95 |

Note the comparison is only for the BMA, where we have both datasets (Miyun and Yanqing counties not included).

We overlaid residential parcels generated by our approach and BICP, and the overlapping area is 211.5 km$^2$ (56.3% out of total 375.6km$^2$ OSM-based residential parcels in Beijing City). In other words, the parcel level validation suggests that, despite only using online open data, our OSM-based approach could conduct reasonably good approximations of data produced by and the conventional manual method.

*5.2 Validating population density with buildings*

We computed the total floor space for each identified residential parcel using available floor space information in 2010. Here, the floor space is used as a proxy of development density and population density. The Pearson correlation coefficient between floor spaces and inferred population numbers is 0.858 for all residential parcels, suggesting that ubiquitously available POI data could be used as a proxy for population density.

*5.3 Validating population attributes*

More details on validating population attributes are available in the study by Long and Shen (2013), in which we validated our proposed Agenter model with a household travel survey with 208,291 individuals. The average similarity index of Agenter is 72.6%, which is significantly greater than that of the null model (43.9%), indicating that Agenter generates sounder synthesized urban residents.

---

[4] Parcels by ORDNANCE in Beijing were similar with those by planners in BICP in terms of parcel size.

## 6 Conclusions

Aiming at the paucity of fine-scale population density and their attributes in cities of the developing world, our study proposes a novel and scalable empirical framework for automatic population spatialization and synthesis using ubiquitously available OSM, POIs and surveys. Consecutive steps are embedded in the framework. Our analysis represents an attempt to use open data and combine the two procedures (spatialization and synthesis) that are separated in existing literature. Empirical results also suggest that open data could help produce reasonable population data.

The contribution of this paper lies in the following aspects: Firstly, we proposed a robust and straightforward approach to delineating parcels, identifying urban parcels, selecting residential parcels, allocating urban population and synthesizing population attributes. Secondly, we employ a novel approach that incorporates a vector-based cellular automata model with the identification of urban parcels, which are associated with a fine spatial scale. Thirdly, considering open data are fast emerging data source, our approach has its potential to be applied to the whole country of China, although here it is developed for Beijing. Our project is also part of the Open Data Initiative, as all our data products will be free and available from the Internet.

The final product of our project is a dataset containing fine scale residents data and their attributes for BMA. This dataset can be applied to but not limited to the following three aspects: first, the parcel-level population density dataset can be referred by urban planners and researchers as an important basic data. For instance, issues like quality-of-life, air pollution exposure and population-based urban agglomeration can be estimated by using the dataset; Secondly, as the dataset we generated contain population distribution and their attributes, it can serve as the direct input, in the form of spatial agents, for emerging agent-based models, which applies coarser dataset as inputs before in the data-sparse environment in China; Lastly, the results have its potential applications in market analysis (e.g. evaluating potential market for retail within the catchment).

With some general limitations of using open data in studying urban dynamics (Liu et al. 2013; Sun et al. 2013), we will conclude with limitations and possible future research avenues that are specific to our population spatialization and synthesis framework. One of the limitations of our approach is that OSM road networks are relatively sparse in many cities (although good enough in our experiment city Beijing) and this will lead to unrealistic large urban parcels (spatial resolution of our study). This should be noted when extending the framework from Beijing to other cities in China or other developing countries. The deficiency of open data is likely to be alleviated by the ever-increasing coverage and quality of OSM data in China. And if possible, more land use/cover datasets should be included in selecting urban parcels out of all delineated parcels. This is expected to increase the precision of selected urban parcels and residential ones.